\documentstyle[preprint,aps,epsfig]{revtex}

\tightenlines

\begin{document}

\title{Chiral symmetry restoration in strange hadronic matter}
\author{P. \  Wang, V. \ E. \ Lyubovitskij,
Th. \ Gutsche, and Amand \, Faessler}
\address{Institut f\"ur Theoretische Physik, Universit\"at
T\"ubingen, Auf der Morgenstelle 14,  \\
D-72076 T\"ubingen, Germany}

\maketitle

\vskip.5cm

\begin{abstract}
The phase transition of chiral symmetry restoration in strange hadronic
matter is studied in the chiral SU(3) quark mean field model. When the
baryon density is larger than a critical density $\rho_c$, the minimal
energy density of the system occurs at the point where the effective
masses of nucleon, $\Lambda$ or $\Xi$ drop to zero. The physical
quantities change discontinuously at this density and the system will be
in the phase of chiral symmetry restoration. A rich phase structure of
strange hadronic matter with different strangeness fraction $f_s$
is obtained.
\end{abstract}

\bigskip

PACS number(s): 21.65.+f; 12.39.-x; 11.30.Rd

\bigskip


\section{Introduction}

It is generally believed that there are two phase transitions in hot
and dense hadronic matter~\cite{Muller,Heinz}. One is deconfinement
which is related to the transition from hadronic to quark matter and
the other is the chiral symmetry restoration. Above the critical
temperature $T_c$, the hadronic states may survive in the high
temperature phase of QCD \cite{Brown2}. In general, deconfinement
has been described in terms of Polyakov loop.
Lattice QCD simulations
show that chiral symmetry restoration and deconfinement occur at the
same critical temperature $T_{c}$~\cite{Karsch}. The qualitative
feature that at high temperature deconfinement and chiral symmetry
restoration come together is not much of a mystery:
in the deconfinement phase quarks are hypothetically free and their
masses tend to zero. It is a longstanding puzzle however why
deconfinement and chiral symmetry restoration occur quantitatively at
the same temperature, though there are some explanations to connect
these two phase transitions~\cite{Wetterich}.

Some of the signals related to the appearance of the deconfined phase
are the phenomenon of strangeness enhancement and the $J/\Psi$
suppression~\cite{Koch,Gonin}. A direct signature of the formation of
the deconfined phase is the existence of strangelets. Many
ultrarelativistic heavy-ion collision experiments at Brookhaven and
CERN are proposed to search for (meta)stable lumps of such kind of
strangelets. Up to now, their is no experiment which confirms the
existence of strangelets. Ardouin et al.~\cite{Ardouin} presented
a novel method which can be applied to characterize the possible
existence of a strange quark matter distillation process in heavy-ion
collisions. Asakawa et al.~\cite{Asakawa} pointed out that the size of
the average fluctuations of net baryon number and electric charge in
a finite volume of hadronic matter differs widely between the confined
and deconfined phases which in turn can be exploited as indicators for
the formation of a quark-gluon plasma in relativistic heavy-ion
collisions. The E864 collaboration found that there was no evidence for
neutral strangelet production in 11.5 GeV/c per nucleon Au+Pb
collisions~\cite{Armstrong}. The new measurement of NA50 revealed
a steady significant decrease in the $J/\Psi$ production rate up to
the most central Pb-Pb collisions and it clearly rules out the
presently available conventional (hadronic) models of $J/\Psi$
suppression. This new observation leads to a natural explanation in
the framework of the formation of a deconfined state of quarks and
gluons~\cite{NA50}.

In the context of the present work we investigate whether chiral
symmetry restoration occurs only when quarks are deconfined, i.e.,
whether the phase transition of chiral symmetry restoration can appear
in hadronic matter. In Ref.~\cite{Bicudo}, it was pointed out that
chiral symmetry breaking will strengthen in nuclear matter.
Furthermore, the quark condensate will increase with baryon density
for values higher than about 0.7$\rho _{0}$, where $\rho _{0}$ is the
saturation density
of nuclear matter. The quark condensate in the nuclear medium has also
been evaluated in the framework of mean-field theory,
the Nambu-Jona-Lasinio model, the relativistic Brueckner Hartree-Fock
approach, and the bare vertex nuclear Schwinger-Dyson
formalism~\cite{Cohen}-\cite{Mitsumori}. The results of these models
indicate that the quark condensate in nuclear matter is considerably
reduced already for saturation density. The effect of chiral symmetry
restoration can also be connected to the effective baryon
mass~\cite{Brown}. At the critical
temperature $T_c$, the hadronic states may survive in the high
temperature phase of QCD \cite{Brown2} where the mesons are massless
in the chiral limit. Due to the Brown-Rho scaling \cite{Brown}, the baryon
mass will also be zero at this time. In general, deconfinement
has been described in terms of Polyakov loop.
Mocsy $et$ $al.$ \cite{Mocsy} illustrated that the deconfinement is
a consequence of chiral symmetry restoration in the presence of
massless quarks in the fundamental representation.
On the other hand, Brown $et$ $al.$ \cite{Brown3} show that chiral
symmetry restoration is accompanied by the vanishing of the trace anomaly
which is connected with the baryon mass. Therefore, the baryon mass can
be used as an order parameter for chiral symmetry restoration.

Presently, the complex action of QCD at finite chemical potential makes it
difficult to study finite-density QCD properties directly from first
principle lattice calculations. We therefore have to resort even more to
phenomenological models to investigate hadronic matter. The symmetries
of QCD can be used as a guideline to determine largely how the hadrons
should interact with each other. Following this idea, models based on
SU(2)$_{L}\times$ SU(2)$_{R}$ symmetry and scale invariance were
proposed. These effective models have been widely used in the last
years to investigate nuclear matter and finite nuclei both at zero and
at finite temperature~\cite{Heide}-\cite{Wang1}. Papazoglou et al.
extended the effective chiral models to SU(3)$_{L}\times$SU(3)$_{R}$
including the full baryon octet~\cite{Papazoglou1,Papazoglou2}.
Although these chiral models indicate that the effective nucleon mass
decreases with density, there is no phase transition with respect to
chiral symmetry restoration. As an extension we proposed the chiral
SU(3) quark mean field model and investigated hadronic and quark
matter~\cite{Wang2}-\cite{Wang4}. The chiral symmetry restoration of
nuclear matter was also studied in this model and a phase transition
was found~\cite{Wang5}. At some critical baryon density, the effective
nucleon mass and the quark condensate decrease discontinuously. In the
present work we will extend the investigation to study chiral symmetry
restoration of strange hadronic matter which includes $\Lambda$,
$\Sigma$ and $\Xi$ hyperons.

The paper is organized as follows. The basic features of the model are
introduced in Sec. II. In Sec. III, we use this model to investigate
chiral symmetry restoration of strange hadronic matter. In Sec. IV we
present the numerical results. In Sec. V we summarize our conclusions.

\section{The model}

Our considerations are based on the chiral SU(3) quark mean field model
(for details see Refs.~\cite{Wang2}-\cite{Wang3}), which contains
quarks and mesons as basic degrees of freedom. Quarks are confined
into baryons by an effective potential. The quark meson interaction
and meson self-interaction are based on the SU(3) chiral symmetry.
Through the mechanism of spontaneous chiral symmetry breaking,
the resulting
constituent quarks and mesons (except for the pseudoscalar ones) obtain
masses. The introduction of an explicit symmetry breaking term in the
meson self-interaction generates the masses of the pseudoscalar mesons
which satisfy the PCAC relation. The explicit symmetry breaking term of
the quark meson interaction leads in turn to reasonable hyperon
potentials in hadronic matter. For completeness, we introduce the main
concepts of the model in this section.

In the chiral limit, the quark field $q$ can be split into left and
right-handed parts $q_{L}$ and $q_{R}$: $q\,=\,q_{L}\,+\,q_{R}$.
Under SU(3)$_{L}\times$ SU(3)$_{R}$ they transform as
\begin{equation}
q_{L} \rightarrow q_{L}^{\prime }\,=\,L\,q_{L},~~~~~
q_{R} \rightarrow q_{R}^{\prime }\,=\,R\,q_{R}\,.
\end{equation}
The spin-0 mesons are written in the compact form
\begin{equation}
M(M^{+})=\Sigma \pm i\Pi =\frac{1}{\sqrt{2}}\sum_{a=0}^{8}
\left( s^{a}\pm i p ^{a}\right) \lambda ^{a},
\end{equation}
where $s^{a}$ and $p ^{a}$ are the nonets of scalar and pseudoscalar
mesons, respectively, $\lambda ^{a}(a=1,...,8)$ are the Gell-Mann
matrices, and $\lambda ^{0}=\sqrt{\frac{2}{3}}\,I$. The alternatives,
plus and minus signs correspond to $M$ and $M^{+}$. Under chiral SU(3)
transformations, $M$ and $M^{+}$ transform as
$M\rightarrow M^{\prime }=LMR^{+}$ and $M^{+}\rightarrow
M^{+^{\prime }}=RM^{+}L^{+}$. As for the spin-0 mesons, the spin-1
mesons are set up in a similar way as
\begin{equation}
l_{\mu }(r_{\mu })=\frac{1}{2}\left( V_{\mu }\pm A_{\mu }\right)
= \frac{1}{2\sqrt{2}}\sum_{a=0}^{8}\left( v_{\mu }^{a}\pm a_{\mu }^{a}
\right) \lambda^{a}
\end{equation}
with the transformation properties:
$l_{\mu}\rightarrow l_{\mu }^{\prime }=Ll_{\mu }L^{+}$,
$r_{\mu}\rightarrow r_{\mu }^{\prime }=Rr_{\mu }R^{+}$.
The matrices $\Sigma$, $\Pi$, $V_{\mu }$ and $A_{\mu }$ can be
written in a form where the physical states are explicit.
For the scalar and vector nonets, we have the expressions
\begin{eqnarray}
\Sigma = \frac1{\sqrt{2}}\sum_{a=0}^8 s^a \, \lambda^a=\left(
\begin{array}{lcr}
\frac1{\sqrt{2}}\left(\sigma+a_0^0\right) & a_0^+ & K^{*+} \\
a_0^- & \frac1{\sqrt{2}}\left(\sigma-a_0^0\right) & K^{*0} \\
K^{*-} & \bar{K}^{*0} & \zeta
\end{array}
\right),
\end{eqnarray}
\begin{eqnarray}
V_\mu = \frac1{\sqrt{2}}\sum_{a=0}^8 v_\mu^a \, \lambda^a=\left(
\begin{array}{lcr}
\frac1{\sqrt{2}}\left(\omega_\mu+\rho_\mu^0\right)
& \rho_\mu^+ & K_\mu^{*+}\\
\rho_\mu^- & \frac1{\sqrt{2}}\left(\omega_\mu-\rho_\mu^0\right)
& K_\mu^{*0}\\
K_\mu^{*-} & \bar{K}_\mu^{*0} & \phi_\mu
\end{array}
\right).
\end{eqnarray}
Pseudoscalar and pseudovector nonet mesons can be written in
a similar fashion.

The total effective Lagrangian is set up as
\begin{eqnarray}
{\cal L}_{{\rm eff}} \, = \, {\cal L}_{q0} \, + \, {\cal L}_{qM}
\, + \,
{\cal L}_{\Sigma\Sigma} \,+\, {\cal L}_{VV} \,+\, {\cal L}_{\chi SB}\,
+ \, {\cal L}_{\Delta m_s} \, + \, {\cal L}_{h}, + \, {\cal L}_{c},
\end{eqnarray}
where ${\cal L}_{q0} =\bar q \, i\gamma^\mu \partial_\mu \, q$ is the
free part for massless quarks. The quark-meson interaction
${\cal L}_{qM}$ can be written in a chiral SU(3) invariant way as
\begin{eqnarray}
{\cal L}_{qM}=g_s\left(\bar{\Psi}_LM\Psi_R+\bar{\Psi}_RM^+\Psi_L\right)
-g_v\left(\bar{\Psi}_L\gamma^\mu l_\mu\Psi_L+\bar{\Psi}_R\gamma^\mu
r_\mu\Psi_R\right)~~~~~~~~~~~~~~~~~~~~~~~  \nonumber \\
=\frac{g_s}{\sqrt{2}}\bar{\Psi}\left(\sum_{a=0}^8 s_a\lambda_a
+ i \gamma^5 \sum_{a=0}^8 p_a\lambda_a
\right)\Psi -\frac{g_v}{2\sqrt{2}}
\bar{\Psi}\left( \gamma^\mu \sum_{a=0}^8
 v_\mu^a\lambda_a
- \gamma^\mu\gamma^5 \sum_{a=0}^8
a_\mu^a\lambda_a\right)\Psi.
\end{eqnarray}
The chiral invariant quark-meson interaction can be traced back to
the Georgi-Manohar model \cite{Manohar} where the chiral symmetry is realized in
nonlinear way with pseudoscalar mesons. In our model, we
include both the scalar and vector mesons which are important in
describing hadronic matter. The constituent quark masses are obtained
through the spontaneous breaking of symmetry and they are not added
by hand as in the Georgi-Manohar model.
In the mean field approximation, the chiral-invariant scalar meson
${\cal L}_{\Sigma\Sigma}$ and vector meson ${\cal L}_{VV}$
self-interaction terms are written as~\cite{Wang2}-\cite{Wang3}
\begin{eqnarray}
{\cal L}_{\Sigma\Sigma} &=& -\frac{1}{2} \, k_0\chi^2
\left(\sigma^2+\zeta^2\right)+k_1 \left(\sigma^2+\zeta^2\right)^2
+k_2\left(\frac{\sigma^4}2 +\zeta^4\right)+k_3\chi\sigma^2\zeta
\nonumber \\ \label{scalar}
&&-k_4\chi^4-\frac14\chi^4 {\rm ln}\frac{\chi^4}{\chi_0^4} +
\frac{\delta}
3\chi^4 {\rm ln}\frac{\sigma^2\zeta}{\sigma_0^2\zeta_0}, \\
{\cal L}_{VV}&=&\frac{1}{2} \, \frac{\chi^2}{\chi_0^2} \left(
m_\omega^2\omega^2+m_\rho^2\rho^2+m_\phi^2\phi^2\right)+g_4
\left(\omega^4+6\omega^2\rho^2+\rho^4+2\phi^4\right), \label{vector}
\end{eqnarray}
where $\delta = 6/33$; $\sigma_0$, $\zeta_0$ and $\chi_0$ are the
vacuum expectation values~\cite{Wang4} of the corresponding mean fields
$\sigma$, $\zeta$ and $\chi$.

Chiral symmetry requires
the following basic relations for the quark-meson coupling constants:
\begin{eqnarray}
\frac{g_s}{\sqrt{2}}
&=& g_{a_0}^u = -g_{a_0}^d = g_\sigma^u = g_\sigma^d = \ldots =
\frac{1}{\sqrt{2}}g_\zeta^s,
~~~~~g_{a_0}^s = g_\sigma^s = g_\zeta^u = g_\zeta^d = 0 \, ,\\
\frac{g_v}{2\sqrt{2}}
&=& g_{\rho^0}^u = -g_{\rho^0}^d = g_\omega^u = g_\omega^d = \ldots =
\frac{1}{\sqrt{2}}g_\phi^s,
~~~~~g_\omega^s = g_{\rho^0}^s = g_\phi^u = g_\phi^d = 0 .
\end{eqnarray}
Note, the values of $\sigma_0$, $\zeta_0$ and $\chi_0$ are determined
from a minimization of the thermodynamical potential~\cite{Wang4}
(see below the procedure for $\sigma_0$ and $\zeta_0$
in Eqs.~(\ref{eq_sigma}) and (\ref{eq_zeta})). On the other hand,
the parameters $\sigma_0$ and $\zeta_0$ are constrained by the
spontaneous breaking of chiral symmetry and are expressed by
the pion ($F_\pi$ = 93~MeV) and the kaon ($F_K$ = 115~MeV)
leptonic decay constants as:
\begin{eqnarray}\label{sigma_0}
\sigma_0 = - F_\pi    \hspace*{1cm} \hspace*{1cm}
\zeta_0  = \frac{1}{\sqrt{2}} ( F_\pi - 2 F_K)
\end{eqnarray}
Masses of vector mesons derived in Eq.~(\ref{vector}) are density
dependent and are expressed as
\begin{eqnarray}
m^2_\omega = m_\rho^2 =
\frac{m_v^2}{1 - \frac{1}{2} \mu \sigma^2_0}\, ,
\hspace*{.5cm} {\rm and} \hspace*{.5cm}
m^2_\phi = \frac{m_v^2}{1 - \mu \zeta^2_0}\, ,
\end{eqnarray}
where the vacuum value of the vector meson mass $m_v=673.6$~MeV
and the density parameter $\mu=2.34$ fm$^2$ are chosen to
reproduce $m_\omega=783$~MeV and $m_\phi=1020$~MeV.

The Lagrangian ${\cal L}_{\chi SB}$ generates the
nonvanishing masses of pseudoscalar mesons
\begin{equation}\label{L_SB}
{\cal L}_{\chi SB}=\frac{\chi^2}{\chi_0^2}\left[m_\pi^2F_\pi\sigma +
\left(
\sqrt{2} \, m_K^2F_K-\frac{m_\pi^2}{\sqrt{2}} F_\pi\right)\zeta\right],
\end{equation}
leading to a nonvanishing divergence of the axial currents which in
turn satisfy the partial conserved axial-vector current (PCAC)
relations for $\pi$ and $K$ mesons. Pseudoscalar,
scalar mesons and also the dilaton field $\chi$ obtain mass terms by
spontaneous breaking of chiral symmetry in the Lagrangian
(\ref{scalar}). The masses of $u$, $d$ and $s$ quarks are generated by
the vacuum expectation values of the two scalar mesons $\sigma$ and
$\zeta$. To obtain the correct constituent mass of the strange quark,
an additional mass term has to be added:
\begin{eqnarray}
{\cal L}_{\Delta m_s} = - \Delta m_s \bar q S q
\end{eqnarray}
where $S \, = \, \frac{1}{3} \, \left(I - \lambda_8\sqrt{3}\right) =
{\rm diag}(0,0,1)$ is the strangeness quark matrix. Based on above
mechanisms, the quark constituent masses are finally given by
\begin{eqnarray}
m_u=m_d=-\frac{g_s}{\sqrt{2}}\sigma_0
\hspace*{.5cm} \mbox{and} \hspace*{.5cm}
m_s=-g_s \zeta_0 + \Delta m_s.
\end{eqnarray}
The parameters $g_s = 4.76$ and $\Delta m_s = 29$~MeV are determined
from $m_q=313$~MeV and $m_s=490$~MeV.
In order to obtain reasonable hyperon potentials in hadronic matter we
include an additional coupling between strange quarks and the scalar
mesons $\sigma$ and $\zeta$~\cite{Wang2}. This term is expressed as
\begin{eqnarray}
{\cal L}_h \,=\, (h_1 \, \sigma \, + \, h_2 \, \zeta) \, \bar{s} s \,.
\end{eqnarray}
In the quark mean field model, quarks are confined in baryons
by the Lagrangian ${\cal L}_c=-\bar{\Psi} \, \chi_c \, \Psi$.
The Dirac equation for a quark field $\Psi_{ij}$ under the additional
influence of the meson mean fields is given by
\begin{equation}
\left[-i\vec{\alpha}\cdot\vec{\nabla}+\chi_c(r)+\beta m_i^*\right]
\Psi_{ij}=e_i^*\Psi_{ij}, \label{Dirac}
\end{equation}
where $\vec{\alpha} = \gamma^0 \vec{\gamma}$\,, $\beta = \gamma^0$\,,
the subscripts $i$ and $j$ denote the quark $i$ ($i=u, d, s$)
in a baryon of type $j$ ($j=N, \Lambda, \Sigma, \Xi$)\,;
$\chi_c(r)$ is a confinement potential, i.e. a static potential
providing confinement of quarks by meson mean-field configurations.

The quark mass $m_i^*$ and energy $e_i^*$ are defined as
\begin{equation}
m_i^*=-g_\sigma^i\sigma - g_\zeta^i\zeta+m_{i0}
\end{equation}
and
\begin{equation}
e_i^*=e_i-g_\omega^i\omega-g_\phi^i\phi \,,
\end{equation}
where $e_i$ is the energy of the quark under the influence of
the meson mean fields. Here $m_{i0} = 0$ for $i=u,d$ (nonstrange quark)
and $m_{i0} = \Delta m_s = 29$~MeV for $i=s$ (strange quark).

The confining potential $\chi_{c}$ is chosen as a combination of scalar
(S) and scalar-vector (SV) potentials as in Ref.~\cite{Wang3,Wang5}:
\begin{eqnarray}
\chi_{c}(r)=\frac12 [\,\chi_{c}^{\rm S}(r)
                         + \chi_{c}^{\rm SV}(r)\,]
\end{eqnarray}
with
\begin{eqnarray}
\chi_{c}^{\rm S}(r)=\frac14 k_{c} \, r^2 \,,
\end{eqnarray}
and
\begin{eqnarray}
\chi_{c}^{\rm SV}(r)=\frac14 k_{c} \, r^2(1+\gamma^0) \,.
\end{eqnarray}
The coupling $k_{c}$ is taken as
$k_{c} = 1$ (GeV $\times$ fm$^{-2})$
to get baryon radii of about 0.6 fm. Using the solution of the Dirac
equation~(\ref{Dirac}) for the quark energy $e_i^*$ we can define
the effective mass of the baryon $j$ as
\begin{eqnarray}
M_j^*=\sqrt{E_j^{*2}- <p_{j \, cm}^{*2}>}\,,
\end{eqnarray}
where $E_j^*=\sum_in_{ij}e_i^*+E_{j \, spin}$ is the baryon energy and
$<p_{j \, cm}^{*2}>$ is the subtraction of the spurious center of mass
motion. In the expression for the baryon energy $n_{ij}$ is the number
of quarks with flavor $"i"$ in a baryon
with flavor $j$ with $j = N \, \{p, n\}\,,
\Sigma \, \{\Sigma^\pm, \Sigma^0\}\,, \Xi \,\{\Xi^0, \Xi^-\}\,,
\Lambda\,$  and $E_{j \, spin}$ is the correction
to the baryon energy due to the spin-spin quark interactions.
We determine the values of $E_{j \, spin}$ as
\begin{eqnarray}
E_{N \, spin} = - 770.3 \,\,\, {\rm MeV}\,,
\hspace*{.5cm}
E_{\Lambda \, spin} = - 756.9 \,\,\, {\rm MeV}\,, \nonumber\\
&&\\
E_{\Sigma \, spin} =  - 690.9 \,\, {\rm MeV}\,, \hspace*{.5cm}
E_{\Xi \, spin} =  - 717.6 \,\, {\rm MeV}\, \nonumber
\end{eqnarray}
from a fit to the data for baryon masses: $M_N = 939$ MeV,
$M_\Lambda = 1116$ MeV,  $M_\Sigma = 1196$ MeV
and $M_\Xi = 1318$ MeV. For illustration we also present the vacuum
values of quark energies and of the center of mass corrections:
\begin{eqnarray}
&&e_u = e_d = 663.0 \,\,\, {\rm MeV}\,, \hspace*{.5cm}
e_s = 799.2 \,\,\, {\rm MeV}\,,\\
&&\nonumber\\
&&<p_{N \, cm}^{2}> = 15.5 \,\,\, {\rm fm}^{-2}\,, \hspace*{.25cm}
<p_{\Lambda \, cm}^{2}> = <p_{\Sigma \, cm}^{2}> =
16.1 \,\,\, {\rm fm}^{-2}\,, \hspace*{.25cm}
<p_{\Xi \, cm}^{2}> = 16.6 \,\,\, {\rm fm}^{-2}\,.  \nonumber
\end{eqnarray}

\section{strange hadronic matter}

Based on the previously defined quark mean field model
the effective Lagrangian for study of strange hadronic
systems is written as
\begin{eqnarray}
{\cal L}&=&\sum\limits_{B = N, \Lambda, \Sigma, \Xi}
\bar{\psi}_B \, \gamma^\mu \, ( \, i \, \partial_\mu \, - \,
M_B^* \, - \, g_\omega^B \, \omega_\mu
\, - \, g_\phi^B \, \phi_\mu\,) \, \psi_B\nonumber \\
&+&\frac12\partial_\mu\sigma\partial^\mu\sigma\, + \,
\frac12\partial_\mu\zeta\partial^\mu\zeta
\, + \, \frac12\partial_\mu \chi\partial^\mu\chi
\, - \, \frac14F_{\mu\nu}F^{\mu\nu} \, - \,
\frac14S_{\mu\nu}S^{\mu\nu}\, + \, {\cal L}_M,
\label{bmeson}
\end{eqnarray}
where
\begin{equation}
F_{\mu\nu}=\partial_\mu\omega_\nu-\partial_\nu\omega_\mu
\hspace*{.5cm} \mbox{and} \hspace*{.5cm}
S_{\mu\nu}=\partial_\mu\phi_\nu-\partial_\nu\phi_\mu
\end{equation}
are the conventional vector meson field ($\omega$ and $\phi$)
strength tensors.

The term
\begin{equation}
{\cal L}_M = {\cal L}_{\Sigma\Sigma} + {\cal L}_{VV}
+ {\cal L}_{\chi SB}
\end{equation}
describes the interaction between mesons which
includes the scalar meson self-interaction ${\cal L}_{\Sigma\Sigma}$,
the vector meson self-interaction ${\cal L}_{VV}$ and the explicit
chiral symmetry breaking term ${\cal L}_{\chi SB}$ defined
previously in Eqs.~(\ref{scalar}), (\ref{vector}) and (\ref{L_SB}).
The Lagrangian ${\cal L}_M$ involves scalar ($\sigma$, $\zeta$
and $\chi$) and vector ($\omega$ and $\phi$) mesons.
The interactions
between quarks and scalar mesons result in the effective baryon masses
$M_B^*$, where subscript $B$ labels the baryon flavor
$B = N, \Lambda, \Sigma$ or $\Xi$.
The interactions between quarks and vector mesons generate the
baryon-vector meson interaction terms of Eq.~(\ref{bmeson}). The
corresponding vector coupling constants $g_\omega^B$ and $g_\phi^B$
satisfy the SU(3) flavor symmetry relations:
\begin{equation}
g_\omega^\Lambda=g_\omega^\Sigma=2g_\omega^\Xi=\frac23g_\omega^N
=2g_\omega^u=\frac{g_v}{\sqrt{2}}
\hspace*{.5cm} \mbox{and} \hspace*{.5cm}
g_\phi^\Lambda=g_\phi^\Sigma=\frac12g_\phi^\Xi=\frac{\sqrt{2}}
3g_\omega^N=g_\phi^s=\frac{g_v}{2} .
\end{equation}
At finite temperature and density, the thermodynamical potential is
defined as
\begin{eqnarray}
\Omega &=& - \sum_{B = N\,, \Lambda\,, \Sigma\,, \Xi }
\frac{g_B k_{B}T}
{(2\pi)^3}\int_0^\infty d^3k\biggl\{{\rm ln}
\left( 1+e^{- [ E_B^{\ast}(k) - \nu_B ]/k_{B}T}\right) \\
&+& {\rm ln}\left( 1+e^{- [ E_B^{\ast}(k)+\nu_B ]/k_{B}T}
\right) \biggr\} -{\cal L}_{M},  \nonumber
\end{eqnarray}
where $E_B^{\ast }(k)=\sqrt{M_B^{\ast 2}+k^{2}}$
and $g_B$ is the degeneracy of baryon $B$ ($g_{N, \Xi}=2$,
$g_\Lambda=1$ and $g_\Sigma=3$). The quantity $\nu _B$
is related to the usual chemical potential $\mu_B$ by
\begin{eqnarray}
\nu_B = \mu_B - g_{\omega}^B\omega - g_{\phi}^B\phi \,.
\end{eqnarray}
At zero temperature, the thermodynamical potential of symmetric
hadronic matter can be expressed as
\begin{eqnarray}
\Omega &=&\sum_{B = N\,, \Lambda\,, \Sigma\,, \Xi }
\frac{g_B}{12\pi^{2}}
\biggl\{ \nu_B\left[ \nu_B^{2} - M_B^{\ast 2}\right]^{1/2}
\left[ 2\nu_B^{2} - 5M_B^{\ast 2}\right] \\
&+& 3M_B^{\ast 4}
{\rm ln}\biggl[\frac{\nu_B + (\nu_B^{2} - M_B^{\ast 2})^{1/2}}
{M_B^{\ast }}\biggr] \biggr\} -{\cal L}_{M}. \nonumber
\end{eqnarray}
The energy per volume and the pressure of the system can be derived as
$\varepsilon =\Omega + \nu_B\rho_B$ and $p = - \Omega$, respectively,
where $\rho_B$ is the baryon density.

The mean field equations for the mesons $\phi _{i}$ are obtained with
$\frac{\partial \Omega}{\partial \phi_i}=0$. For example,
the equations for the scalar mesons $\sigma$, $\zeta$ are
expressed as
\begin{eqnarray}\label{eq_sigma}
k_{0}\chi^{2}\sigma
&-&4k_{1}\left( \sigma^{2} \, + \, \zeta^{2}\right) \sigma \, - \,
2k_{2} \sigma^{3}\, - \, 2k_{3}\chi \sigma \zeta \, - \,
\frac{2\delta }{3\sigma }\chi^{4}
\, + \, \frac{\chi^{2}}{\chi _{0}^{2}}m_{\pi }^{2}F_{\pi } \nonumber \\
&-&\left( \frac{\chi }{\chi _{0}}\right)^{2}m_{\omega }\omega ^{2}
\frac{\partial m_{\omega }}{\partial \sigma }\, + \,
\sum_{B = N\,, \Lambda\,, \Sigma\,, \Xi } \frac{\partial M_{B}^{\ast }}
{\partial \sigma } <\bar{\psi _{B}}\psi_{B}>=0,~~~~~~~~~~
\end{eqnarray}
\begin{eqnarray}\label{eq_zeta}
k_{0}\chi^{2}\zeta &-& 4k_{1}\left(\sigma^{2} \, + \,
\zeta ^{2}\right)
 \zeta \, - \, 4k_{2}\zeta ^{3} \, - \,
k_{3}\chi \sigma ^{2} \, - \, \frac{\delta }{3\zeta }\chi^{4} \, + \,
\frac{\chi^{2}}{\chi _{0}^{2}} \left( \sqrt{2}m_{k}^{2}F_{k} \, - \,
\frac{1}{\sqrt{2}}m_{\pi }^{2}F_{\pi } \right)\nonumber\\
&-&\left( \frac{\chi }{\chi _{0}}\right)^{2}m_{\phi }\phi^{2}
\frac{\partial m_{\phi }}{\partial \zeta } \, + \,
\sum_{B = \Lambda\,, \Sigma\,, \Xi}
\frac{\partial M_{B}^{\ast}}{\partial\zeta}
<\bar{\psi_{B}}\psi_{B}>=0,~~~~~~~~~~~~~~
\end{eqnarray}
where
\begin{eqnarray}
<\bar{\psi_B}\psi_B>&=&\frac{g_B \, M_B^\ast}{\pi^2} \,
\int_{0}^{k_{F_B}} dk \frac{k^2}{\sqrt{M_B^{\ast 2}+k^2}} \\
&=& \frac{g_B \, M_B^{\ast 3}}{2 \, \pi^2} \,
\biggl[ \frac{k_{F_B}}{M_B^\ast} \,
\sqrt{1 + \frac{k_{F_B}^2}{M_B^{\ast 2}}}
 -  {\rm ln}\biggl( \frac{k_{F_B}}{M_B^\ast} +
\sqrt{1 + \frac{k_{F_B}^2}{M_B^{\ast 2}}} \biggr) \biggr]  \nonumber
\end{eqnarray}
is the baryon condensate and
$k_{F_{B}}=\sqrt{\nu_{B}^{2}-M_{B}^{\ast 2}}$ is the Fermi momentum.
At zero baryon density, the vacuum expectation values of $\sigma$
and $\zeta$ have been introduced in Eq.~(\ref{sigma_0}).
With an increase of the baryon density the values of $\sigma$ and
$\zeta$ rise which in turn result in a decrease of the effective baryon
masses. The effective baryon mass $M_B^{\ast}$ are generally in the
region
\begin{eqnarray}
0\leq M_B^{\ast}(\sigma ,\zeta )
\leq M_B^{\ast}(\sigma _{0}, \zeta _{0}) \,.
\end{eqnarray}
The ranges of mean field values for $\sigma$ and $\zeta$
\begin{eqnarray}
\sigma_{0}\leq \sigma \leq \sigma _{b}, \hspace*{.5cm}
\zeta _{0}\leq \zeta \leq \zeta_{b} \,,
\end{eqnarray}
where the upper values $\sigma _{b}$ and
$\zeta _{b}$ are determined by $M_{N}^{\ast }=0$,
$M_{\Lambda }^{\ast }=0$ or $M_{\Xi }^{\ast }=0$. Since the $\Sigma$
hyperon has the same quark flavor content as the $\Lambda$ hyperon,
its effective mass is always larger than that of the $\Lambda$. At a
specific point for high baryon density, the minimum of the
thermodynamical potential $\Omega$ appears not inside the region,
but at the boundary of values given for $\sigma$ and $\zeta$.
In this limit,
Eqs.~(\ref{eq_sigma}) and (\ref{eq_zeta}) for $\sigma$ and
$\zeta$ are not valid any more. At the boundary, the mean fields
$\sigma$ and $\zeta$ satisfy the equations
\begin{equation}
\frac{\partial \Omega }{\partial Z}=0,~~~~~~~dZ=\sqrt{d\sigma ^{2} +
d\zeta^{2}},
\end{equation}
where $Z$ is in the boundary and $d\sigma$, $d\zeta$ are constrained by
\begin{equation}
\frac{\partial M_{B}^{\ast }(\sigma ,\zeta )}{\partial \sigma }
d\sigma + \frac{\partial M_{B}^{\ast }(\sigma ,\zeta )}{\partial\zeta }
d\zeta = 0~~~(B = N\,, \Lambda\,, \Xi).
\end{equation}
Therefore, if the minimum of the thermodynamical potential appears for
$M_{B}^{\ast }(\sigma ,\zeta )=0$, the equations for $\sigma$ and
$\zeta$ can be deduced from
\begin{equation}\label{eq_bound1}
M_{B}^{\ast }(\sigma ,\zeta )=0,
\end{equation}
\begin{equation}\label{eq_bound2}
\frac{\partial \Omega }{\partial \sigma }\frac{1}{\sqrt{1+\left( \frac{
\partial M_{B}^{\ast }/\partial \sigma }{\partial M_{B}^{\ast}/\partial
\zeta }\right) ^{2}}}+\frac{\partial \Omega }{\partial \zeta }
\frac{1}{\sqrt{1+\left( \frac{\partial M_{B}^{\ast }/\partial \zeta }
{\partial M_{B}^{\ast}/\partial \sigma }\right) ^{2}}}=0.
\end{equation}
For the special case, when $M_{N}^{\ast}$ is only a function of
$\sigma$, Eq.~(\ref{eq_bound2}) is identical to Eq.~(\ref{eq_zeta}).

\section{Numerical results}

The parameters of the chiral SU(3) quark mean field model are
determined, as previously done~\cite{Wang2}-\cite{Wang3}, by the meson
masses in vacuum and by the properties of nuclear matter. The confining
potential $\chi_{c}$, is chosen as in Refs.~\cite{Wang3,Wang5},
where it was shown in comparison with two other two types of
potentials that it is the best choice to describe finite systems.
The model parameters are summarized as follows:
\begin{eqnarray}
&&k_0 = 4.21\,, \hspace*{.25cm}
k_1 = 2.26\,, \hspace*{.25cm}
k_2 = -10.16\,, \hspace*{.25cm}
k_3 = -4.38\,, \hspace*{.25cm}
k_4 = -0.13\,, \\
&&g_4 = 7.5\,, \hspace*{.25cm}
h_1 = -2.07\,, \hspace*{.25cm}
h_2 = 2.90\,, \hspace*{.25cm}
g_s = 4.76\,, \hspace*{.25cm}
g_v = 10.37\,, \hspace*{.25cm}
k_c = 1 \,\, ({\rm GeV} \times {\rm fm}^{-2})\,. \nonumber
\end{eqnarray}
We do not mention the parameters which are fixed from experimental
data or from the chiral symmetry constraints.
In the previous section, we already mentioned that the
values of $\sigma$ and $\zeta$ are determined by minimization of the
thermodynamical potential $\Omega$. At zero temperature, the scalar
meson mean field values can be obtained by minimizing the energy per
volume $\varepsilon$. For nonstrange matter, $\varepsilon$ can be
expressed as a function of $\sigma$ or of the effective nucleon mass
$M_{N}^{\ast }$. At some critical baryon density the minimum of
$\varepsilon$ appears at $M_{N}^{\ast}=0$, a scenario which has been
studied in Ref.~\cite{Wang5}. For strange hadronic matter $\varepsilon$
is a function of the mean field values $\sigma$ and $\zeta$.
At a given strangeness fraction, when the baryon density is low, there
is a minimum of $\varepsilon$ where the effective nucleon and
hyperon masses have nonzero values. In most other models, for any
values of the strangeness fraction
and the baryon density, a solution for the mean field equations
corresponding to (\ref{eq_sigma}) and (\ref{eq_zeta}), can be obtained,
that is the minimum of $\varepsilon$ occurs for nonzero effective
baryon masses. However, in the present chiral SU(3) quark mean field
model, the minimum of $\varepsilon$ will appear on the boundary of
$\sigma$ and $\zeta$ at high baryon density for a fixed strangeness
fraction.

In Fig. 1 we plot the boundary of $\sigma$ and $\zeta$ resulting from
the model. The solid, dashed and dotted lines correspond to the three
boundaries where the effective baryon masses are zero, i.e.,
$M_{N}^{\ast }=0$, $M_{\Lambda }^{\ast }=0$ and $M_{\Xi }^{\ast }=0$.
The two coordinate axis and the three boundaries form an area ABCDE.
Inside this area, the effective baryon masses are nonzero. On the
boundary the dynamical equations (\ref{eq_sigma}) and (\ref{eq_zeta})
for $\sigma$ and $\zeta$ have to be replaced by equations
(\ref{eq_bound1}) and (\ref{eq_bound2}). The numerical calculations
show that for small strangeness fraction and at sufficiently high
baryon density the minimum of $\varepsilon$ corresponds to the case of
the solid line with a vanishing effective nucleon mass. For large
strangeness fraction, the minimum of $\varepsilon$ will coincide with
the dotted line defined by a zero effective mass of the $\Xi$ hyperon.
The minimum of $\varepsilon$ will never correspond to case of the
dashed line (except at the points B and C) where only
$M_{\Lambda }^{\ast }$ vanishes. At the point B both $M_{N}^{\ast }$
and $M_{\Lambda }^{\ast }$ are zero, whereas point C is set by
vanishing values of $M_{\Lambda }^{\ast }$ and $M_{\Xi }^{\ast }$.

We next discuss the change of the effective baryon masses in dependence
on the baryon density. In Fig. 2, we plot the effective baryon masses
versus density at a strangeness fraction of $f_{s}=0.8$. As in most
other models, the effective baryon masses decrease with increasing
baryon density. At this value for the strangeness fraction, the
effective mass of the nucleon decreases faster than those of the other
baryons. One the contrary, the effective $\Xi$ mass decreases slower.
This is because when the strangeness is small, the interaction between
non-strange quarks and the $\sigma$ meson is considerably stronger than
the one between strange quarks and $\zeta$. When the baryon density
$\rho_{B}$ reaches a value of 0.46 fm$^{-3}$, the effective nucleon
mass drops from 0.2 GeV to zero. All other effective baryon masses also
decrease discontinuously, but in there cases to finite values. In the
range of 0.46 fm$^{-3}<\rho _{B}<$0.69 fm$^{-3}$ the effective hyperon
masses continue to decrease. When the density is larger than
0.69 fm$^{-3}$, both nucleon and $\Lambda$ masses equal zero while the
$\Sigma$ and $\Xi$ masses remain at a constant and finite value.

The density dependence of the baryon masses sensitively depends on the
strangeness fraction. For high strangeness fraction, the interaction
between hyperons is stronger than the one between nucleons. The
effective masses of hyperons with a larger strange quark content will
therefore decrease faster. As a result, at a high value of the baryon
density, the hyperon masses will be zero, while the nucleon mass is
still finite. This effect is clearly seen from Fig. 3, where we plot
the phase diagram. For the case of nonstrange hadronic matter, there
are only two phases. One normal phase, where chiral symmetry is
spontaneously broken, whereas at high density we have chiral symmetry
restoration phase with a zero effective nucleon mass. Strange hadronic
matter, corresponds to the additional presence of $\Lambda$, $\Sigma$
and $\Xi$ hyperons. Because the effective masses of the baryons drop to
zero at different values of the baryon density and strangeness
fraction, we obtain different chiral symmetry restoration phases. The
rich phase structure is illustrated in Fig. 3 with five qualitatively
different regions. Phase I refers to the normal phase with nonzero
baryon masses.
In the chiral symmetry restoration phase II, only the effective nucleon
mass is zero. In phase III, both nucleon and $\Lambda$ masses are
vanishing. Phase IV, designates the case where both $\Lambda$ and $\Xi$
have zero effective masses, whereas phase V only has a vanishing value
for $M_{\Xi }^{\ast }$. For different value of the strangeness fraction
$f_s$, the normal phase I will turn into different chiral symmetry
restoration phases at different critical densities. For nonstrange
nuclear matter the critical baryon density is reached at a value of
0.28 fm$^{-3}$, where the system will be in phase II. For values of the
strangeness fraction with $f_s<1.0$, the critical density increases
with growing $f_{s}$. The maximum of the critical density of about
0.5 fm$^{-3}$ is reached for $f_{s}\simeq 1.0$. For the range of values
$0.6<f_{s}<1.1$, when the baryon density is high enough, the chiral
restoration phase can furthermore change from phase II to phase III.
For larger values of the strangeness fraction, that is $f_{s}>1.1$, the
effective nucleon mass will not drop to zero at any density. In the
range of $1.1<f_{s}<1.3$, when the density reaches a critical point,
the system changes directly from phase I to phase III. At some higher
density, a further transition from phase III into phase IV takes
places. For $1.3<f_{s}<1.75$, the normal phase I changes into phase IV
at the critical point. For $f_{s}>1.75$, chiral symmetry restoration is
manifest only in phase V.

The energy per baryon E/A is defined as
\begin{equation}
E/A=\frac{\varepsilon}{\rho_B}-\frac{M_N\rho_N+M_\Lambda\rho_\Lambda
+M_\Sigma\rho_\Sigma+M_\Xi\rho_\Xi}{\rho_B}.
\end{equation}
In Fig. 4, we  E/A in dependence on the baryon
density for different strangeness fractions. For normal nuclear matter
the binding energy is 16 MeV at the saturation density $\rho _{0}$ of
0.16 fm$^{-3}$. When the density reaches a value of about
0.28 fm$^{-3}$, a discontinuous decrease of energy per baryon occurs.
At $\rho _{B}\simeq$ 0.31 fm$^{-3}$ there is a second minimum for E/A
which is close to -16 MeV. When the density is larger than about
0.31 fm$^{-3}$ the energy per baryon increases with growing density.
As the strangeness fraction increases, the first minimum of E/A first
increases and then decreases. At $f_{s}\simeq 1.2$, the first maximum
of the binding energy is about 19 MeV. The binding energy at the second
minimum is about 70 MeV and the corresponding $f_{s}$ is about 1.5.
The system therefore favors to have a large strangeness fraction where
the corresponding density is about 3-4 times that of the saturation
density $\rho_{0}$ of normal nuclear matter.

\section{Summary}

We applied the chiral SU(3) quark mean field model to investigate the
mechanism of chiral symmetry restoration in strange hadronic matter.
The model is based on effective quark-meson and meson self-interactions
which satisfy the chiral SU(3) symmetry constraint. Chiral symmetry
restoration of nonstrange nuclear matter was studied in a previous
paper~\cite{Wang5}. When a critical baryon density is reached, the
effective nucleon mass will drop to zero. This phenomenon is related to
the fact that nuclear matter has its lowest energy per volume in the
chiral symmetry restoration phase at high density.

In the present work we extended the discussion to strange hadronic
systems including $\Lambda$, $\Sigma$ and $\Xi$ hyperons. As for
the case of nonstrange nuclear matter, when the baryon density
is sufficient high, the minimum of the energy per volume $\varepsilon$
of strange matter will occur for effective baryon masses reaching
a vanishing value.  As it turns out, the effective masses of the
different baryons do not change to zero at the same value of
baryon density and strangeness fraction. In the effective model
there occur four different chiral symmetry restoration phases
which are characterized by zero effective masses of the different
baryons. For a given strangeness fraction, when the baryon density
is larger than the critical density, hadronic matter will make a
transition to the corresponding chiral symmetry restoration phase
with values for the critical density which is about 2-3 times
larger than the nuclear saturation density $\rho_{0}$.

The energy per baryon E/A has two minima for a given strangeness
fraction. One corresponds to the normal phase, the other one occurs for
the chiral symmetry restoration phase. For small strangeness fraction,
that is $f_s<1.0$, the minimum of E/A in the normal phase is lower than
the one of the restored phase. For large strangeness fraction the
system is favored to be in the chiral symmetry restoration phase since
it occurs with a larger binding energy. The maximal binding energy is
about 70 MeV for a strangeness fraction of about 1.5.

The deconfinement phase transition may also occur at high densities.
It is worthwhile to study whether the phase transition between hadronic
and quark matter also takes place at values for the baryon densities
which are close to the ones when chiral symmetry restoration occurs.
Both the quark and the hadronic phase can be described in the chiral
SU(3) quark mean field model. The model therefore opens up the
possibility to also study the deconfinement phase transition, a topic
which will be pursued in future studies.

\vspace*{1cm}

{\bf Acknowledgments}

\noindent P. W. would like to thank the Institute for Theoretical
Physics, University of T\"ubingen for their hospitality. This work was
supported by the Alexander von Humboldt Foundation and by the DFG under
contracts FA67/25-3, GRK683.

\vfill

\newpage

\centerline{FIGURE CAPTIONS}

\vspace*{.5cm}

Fig. 1. The boundary of $\sigma$ and $\zeta$. The solid, dashed and
dotted lines are the boundaries for $M_N^*=0$, $M_\Lambda^*=0$ and
$M_\Xi^*=0$, respectively.

\bigskip

Fig. 2. The effective baryon masses versus baryon density with
$f_s=0.8$. The solid, dashed, dotted and dash-dotted lines are for
nucleon, $\Lambda$, $\Sigma$ and $\Xi$, respectively.

\bigskip

Fig. 3. The phase diagram for strange hadronic matter. Phase I is the
normal phase. Phase II, III, IV and V are the chiral symmetry
restoration phases with $M_N^*=0$, $M_N^*=M_\Lambda^*=0$,
$M_\Lambda^*=M_\Xi^*=0$ and $M_\Xi^*=0$, respectively.

\bigskip

Fig. 4. The energy per baryon E/A versus baryon density $\rho_B$ at
different strangeness fractions.
\bigskip

\newpage
\begin{figure}
\centering{\
\epsfig{figure=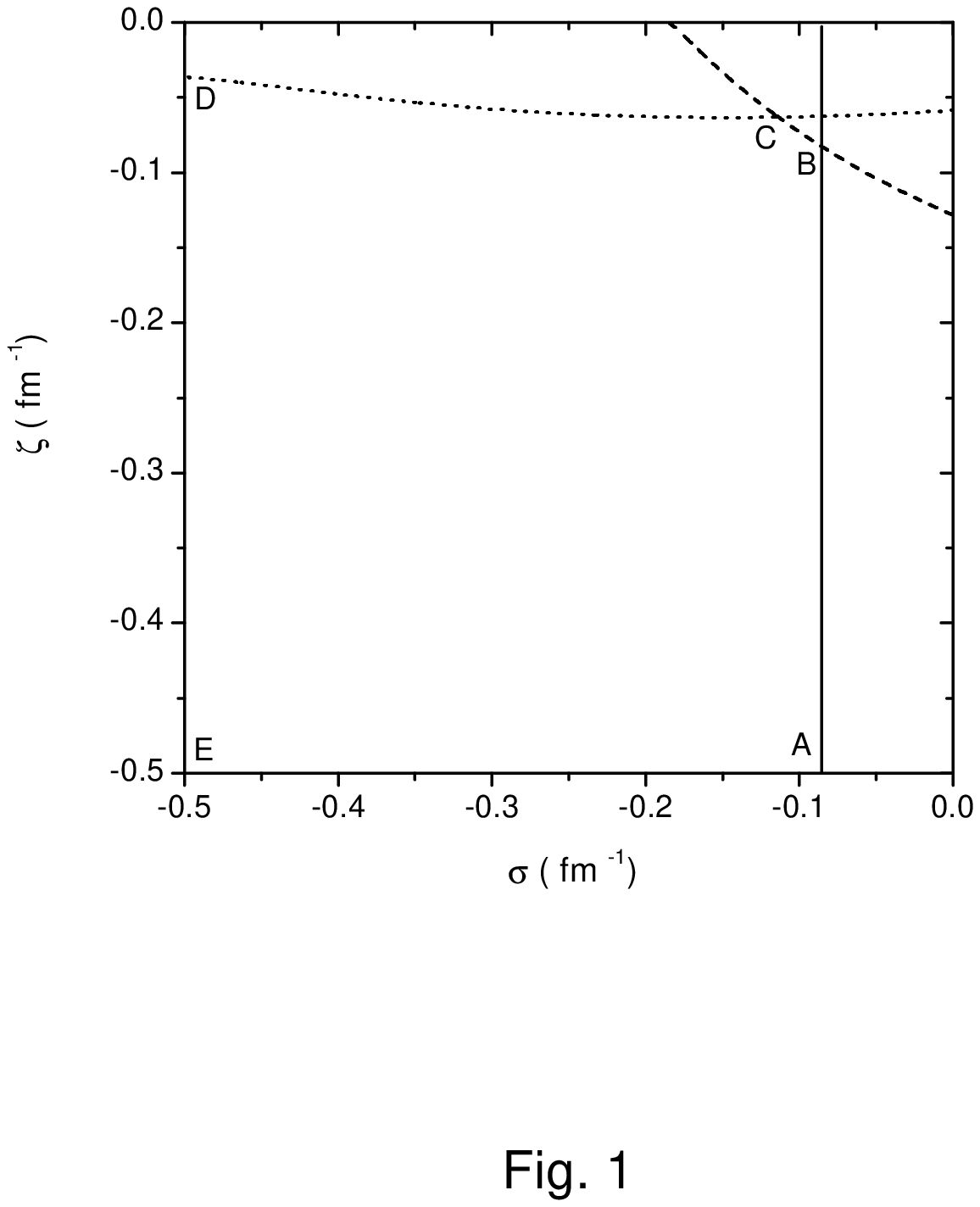,height=12cm}
}
\end{figure}

\newpage
\begin{figure}
\centering{\
\epsfig{figure=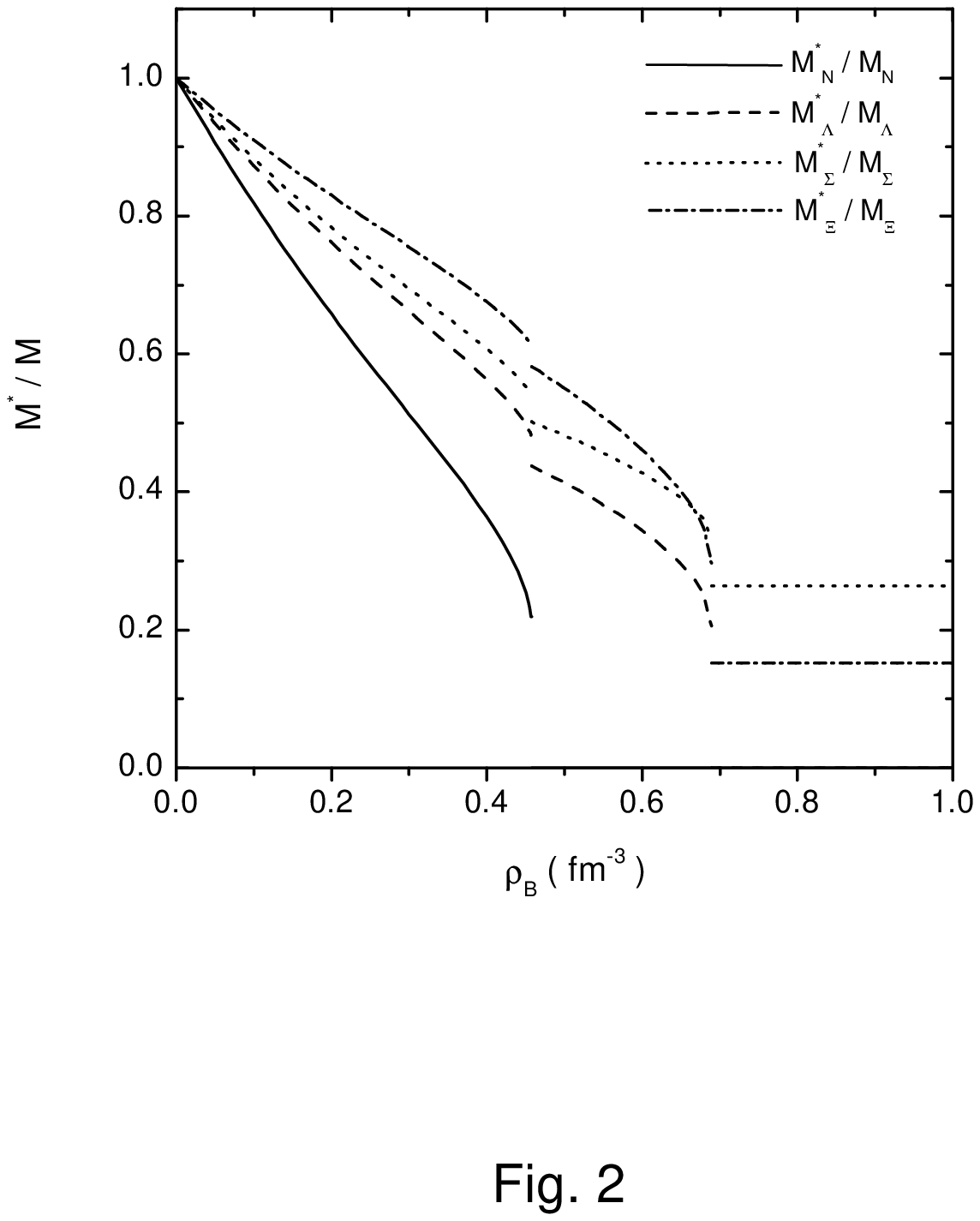,height=12cm}
}
\end{figure}

\newpage
\begin{figure}
\centering{\
\epsfig{figure=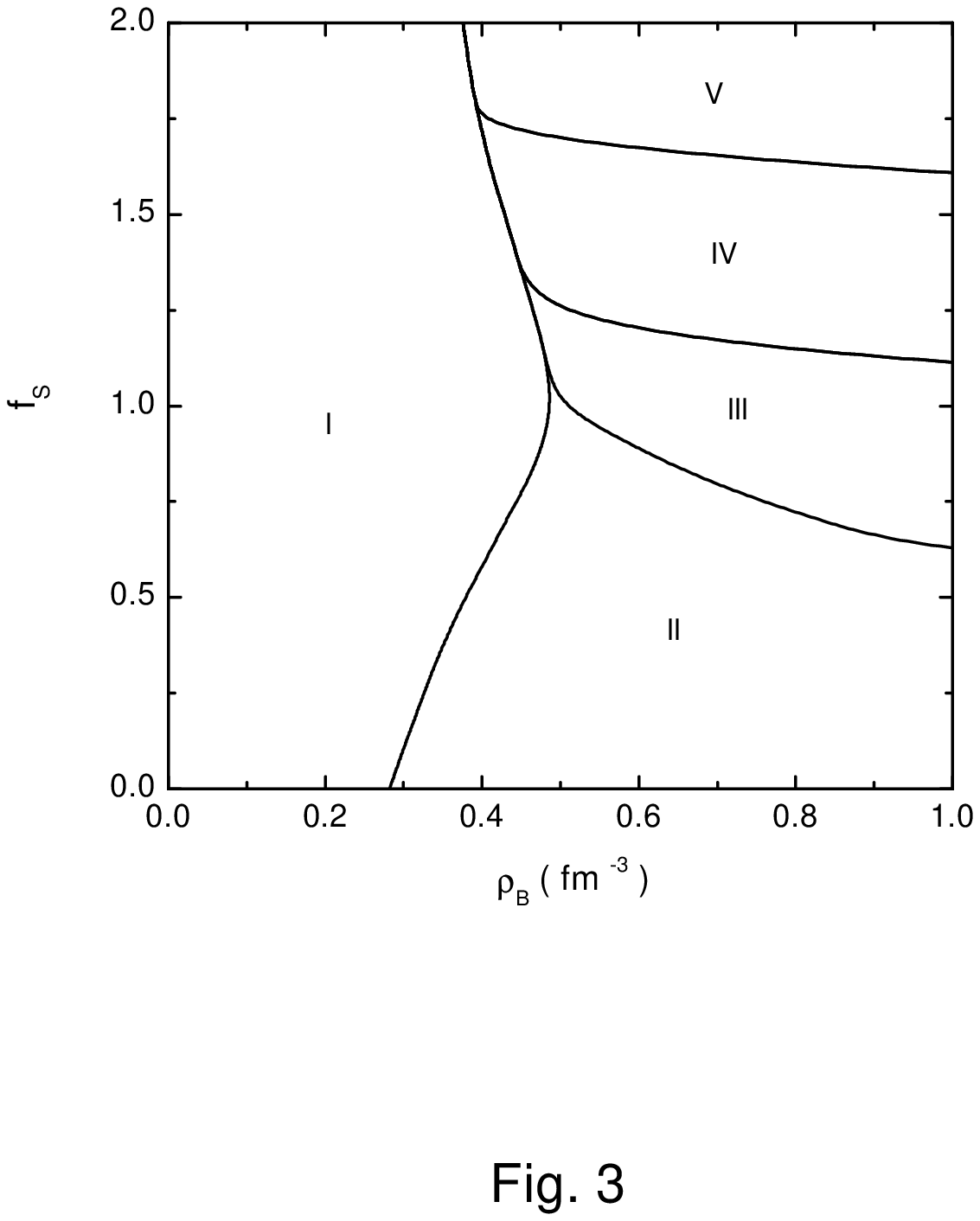,height=12cm}}
\end{figure}

\newpage
\begin{figure}
\centering{\
\epsfig{figure=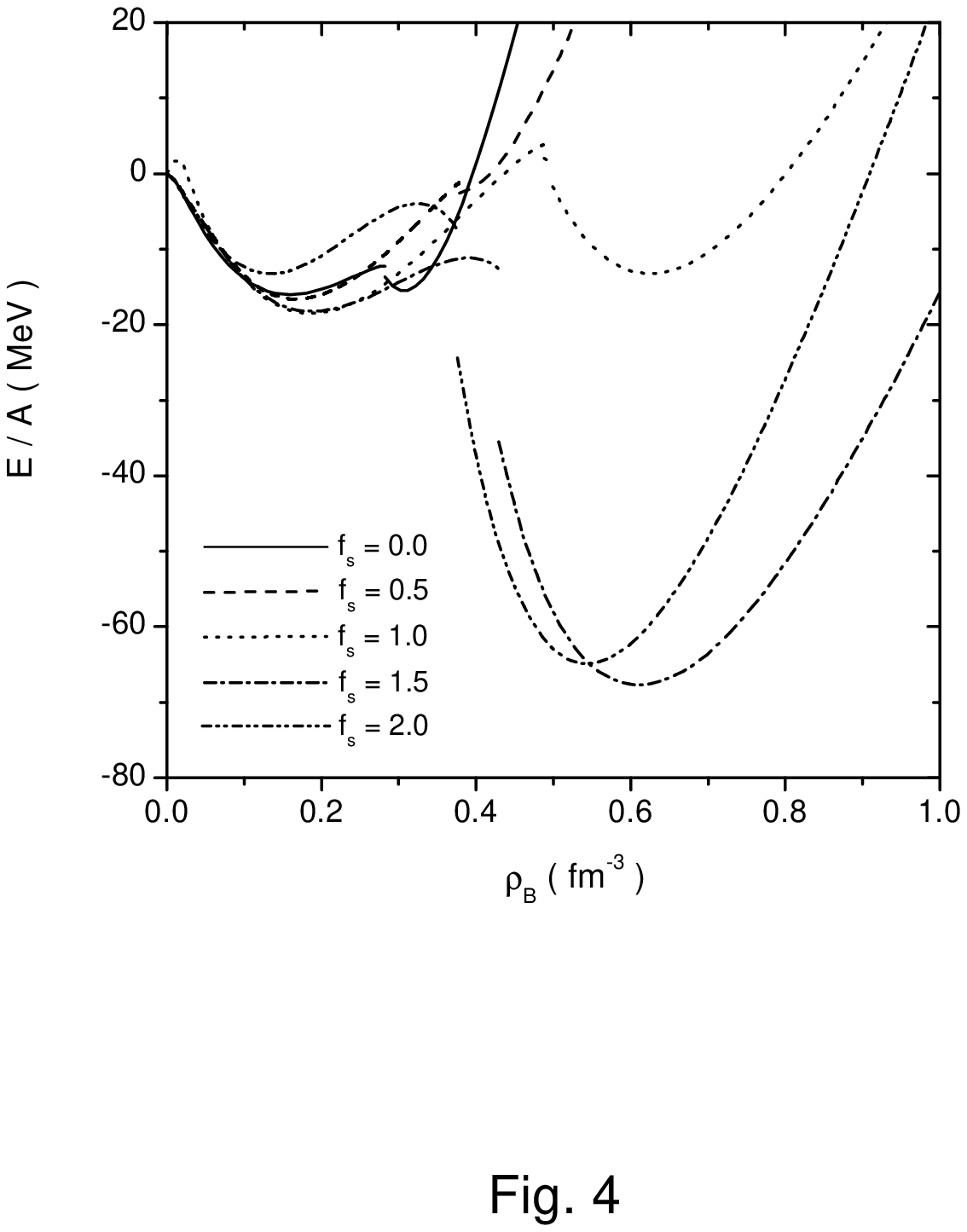,height=12cm}
}
\end{figure}

\end{document}